\documentclass[preprintnumbers,twocolumn,nofootinbib,tightenlines,superscriptaddress]{revtex4}
\usepackage{graphicx}
\usepackage{amssymb}
\usepackage{amsmath,mathrsfs,verbatim}
\usepackage{times}
\usepackage{animate}
\usepackage{latexsym}
\usepackage[usenames]{color}

\begin{document}

\title{Refracting into Ultra Diffuse Galaxy NGC 1052-DF2 by Passing near the Center of NGC 1052}

\author{Ran Huo}
\affiliation{Department of Physics and Astronomy, University of California, Riverside, California 92521, USA}

\date{\today}

\begin{abstract}
The recent observation of the ultra-diffuse galaxy NGC 1052-DF2~\cite{vanDokkum:2018vup} shows a galaxy may lack dark matter. Usually dark matter is much more abundant than stellar in galaxy environment. This dark matter to baryon mass ratio is generally larger than the cosmological ratio of about five, since a significant part of baryon is diffused in the form of intergalactic medium. How to achieve such a low dark matter to baryon mass ratio is a challenge to the standard galaxy formation mechanism. Here we show that such a low ratio can be a natural consequence if the NGC 1052-DF2 had experienced a single passage within a few kpcs to the center of galaxy NGC 1052. The tidal effect of NGC 1052 in the encounter will heat the NGC 1052-DF2 up, stretch the previous dwarf galaxy significantly into its current size. The relative lacking of dark matter in the observed region is a natural consequence of the dark matter extended distribution and relatively less concentration in the corresponding central region before encounter, together with a systematic underestimation of the trace mass estimator method during relaxation after encounter. The observed flat distribution of the ultra-diffuse galaxy can be reproduced. Our results shows no need of introducing any new physical mechanism.
\end{abstract}

\maketitle

The key observation from~\cite{vanDokkum:2018vup} is the Ultra Diffuse Galaxy (UDG) NGC 1052-DF2 with a luminous mass of $2\times10^8~M_\odot$ has a total gravitational mass of less than $3.4\times10^8~M_\odot$ within the central $7.6$~kpc (also see~\cite{2018RNAAS...2b..54V,2018ApJ...863L..15W,2019arXiv190103711D}). Later a second UDG NGC 1052-DF4~\cite{vanDokkum:2019} is found to be very similar. Since then~\cite{2018ApJ...859L...5M,2019MNRAS.484..245L,2018MNRAS.481L..59H,2019MNRAS.484..510N} suggests a weaker constraint on the dark matter (DM) to baryon mass ratio based on analysis of mass estimation uncertainties, but still we consider this is at odds since the natural value for a dwarf galaxy is $M/L\sim10^3$ or so, due to the well established stellar feedback mechanism~\cite{Hopkins:2013vha,2018MNRAS.480..800H} which in particular works efficiently for dwarf galaxies. Later analysis~\cite{2018ApJ...864L..18V,2018RNAAS...2c.146B,2019RNAAS...3b..29V} reassured NGC 1052-DF2(4) to have a distance of about $20$~Mpc and be in association with the luminous elliptical galaxy NGC 1052, which is against~\cite{2018arXiv180610141T}. NGC 1052-DF2 are only visually $14$~arcmin (or $80$~kpc at such distance) apart from the central NGC 1052, which suggests an interesting possibility of interactions between the two such as tidal stripping~\cite{2018MNRAS.480L.106O,2019MNRAS.485..382C,2018arXiv181109070J}.

\begin{figure}[t!]
\includegraphics[height=3in]{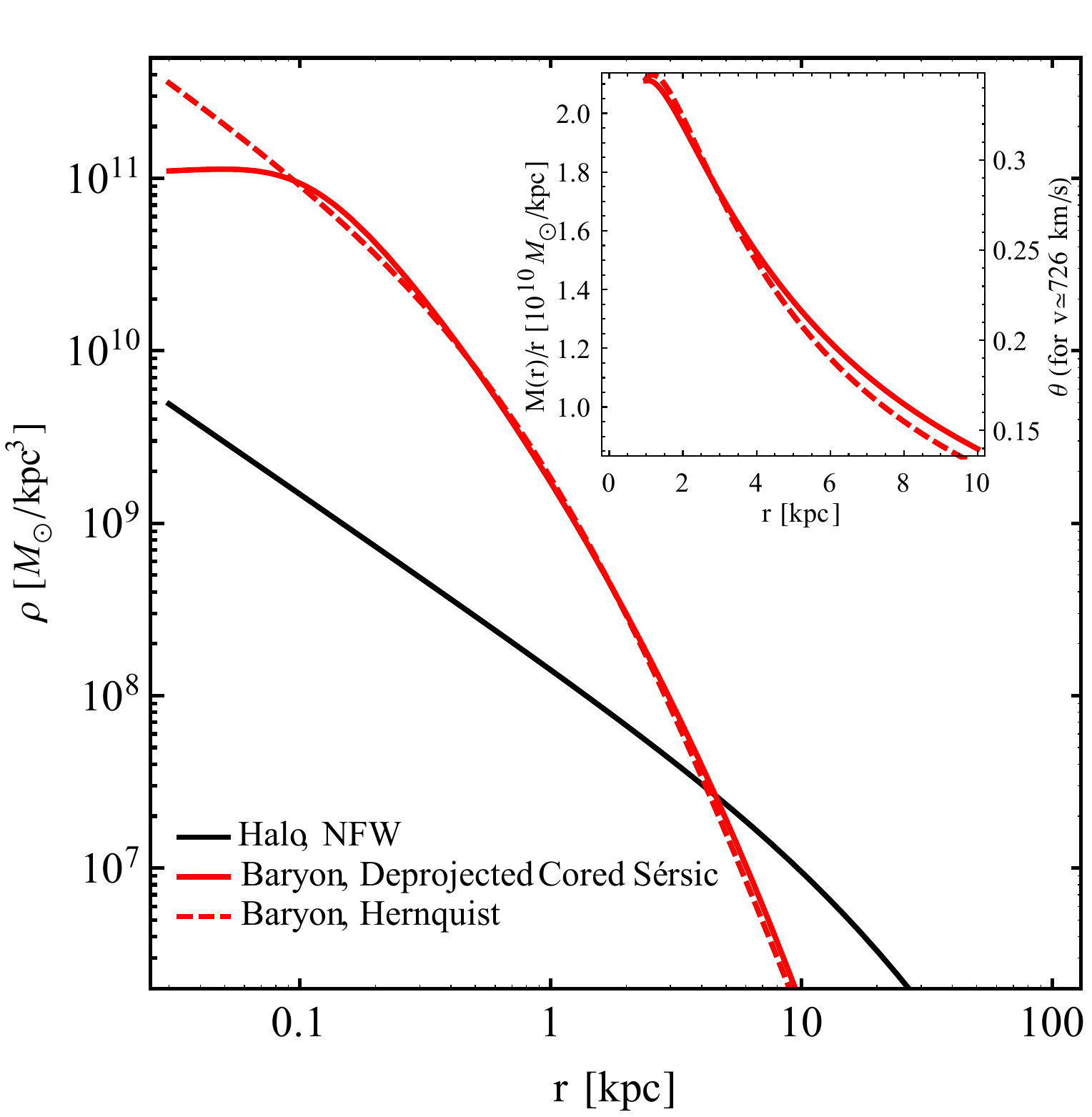}
\caption{\textbf{The simulation used initial density profile for the halo $+$ star system of NGC 1052.} The deprojected cored S\'{e}rsic profile is compared to the widely used Hernquist profile. In the inset we plot the deflection angle with equation~(\ref{eq:deflection}) as a function of pericenter/impact parameter, with effective $v_\infty=v_\text{peri}\simeq726~\text{km}/\text{s}$ extracted from simulation.}
\label{fig:NGC1052}
\end{figure}

Here we will use the NGC 1052-DF2 as an example, and explore the possibility that it has experienced a single but very close encounter with the central part of NGC 1052. NGC 1052 has a cored S\'{e}rsic profile with $M_\ast=10^{11}~M_\odot$, S\'{e}rsic $R_e=21.9$~arcsec or $2.06$~kpc at such distance, and S\'{e}rsic index $n=3.4$~\cite{2017MNRAS.464.4611F}. The profile is reported as cored but there is no reported central deficit core size $R_b$ as well as core parameters $\alpha$ and $\gamma$, here we choose a typical value of $R_b=0.1$~kpc and $\alpha=3$, $\gamma=0$ which should not be essential to the following physics. For S\'{e}rsic index close to $4$ it is well known that the analytical Hernquist profile is a good approximation for the deprojected density profile assuming spherical symmetry, here by matching the 3d half mass radius we determine the radius parameter $r_\text{H}=1.17$~kpc. Such a profile is cuspy and can dominate in density over an NFW profile of halo in the central region. For the DM halo we use $M_\text{halo}=5.54\times10^{12}~M_\odot$ (with $M_\text{vir}=3\times10^{12}~M_\odot$) for the NFW profile, which is the best matched halo corresponding to such $M_\ast$ according to the stellar mass - halo mass relation~\cite{Hopkins:2013vha,2018MNRAS.480..800H}. According to the NFW profile concentration $-$ mass relation~\cite{Dutton:2014xda} the halo has an NFW $r_s=40.6$~kpc initially. In Fig.~\ref{fig:NGC1052} we plot the NGC 1052 halo $+$ star system initial condition (IC), we can see that in the central region it is more baryon dominated, a well known fact due to baryonic dissipation.

\begin{figure}[t!]
\includegraphics[height=3in]{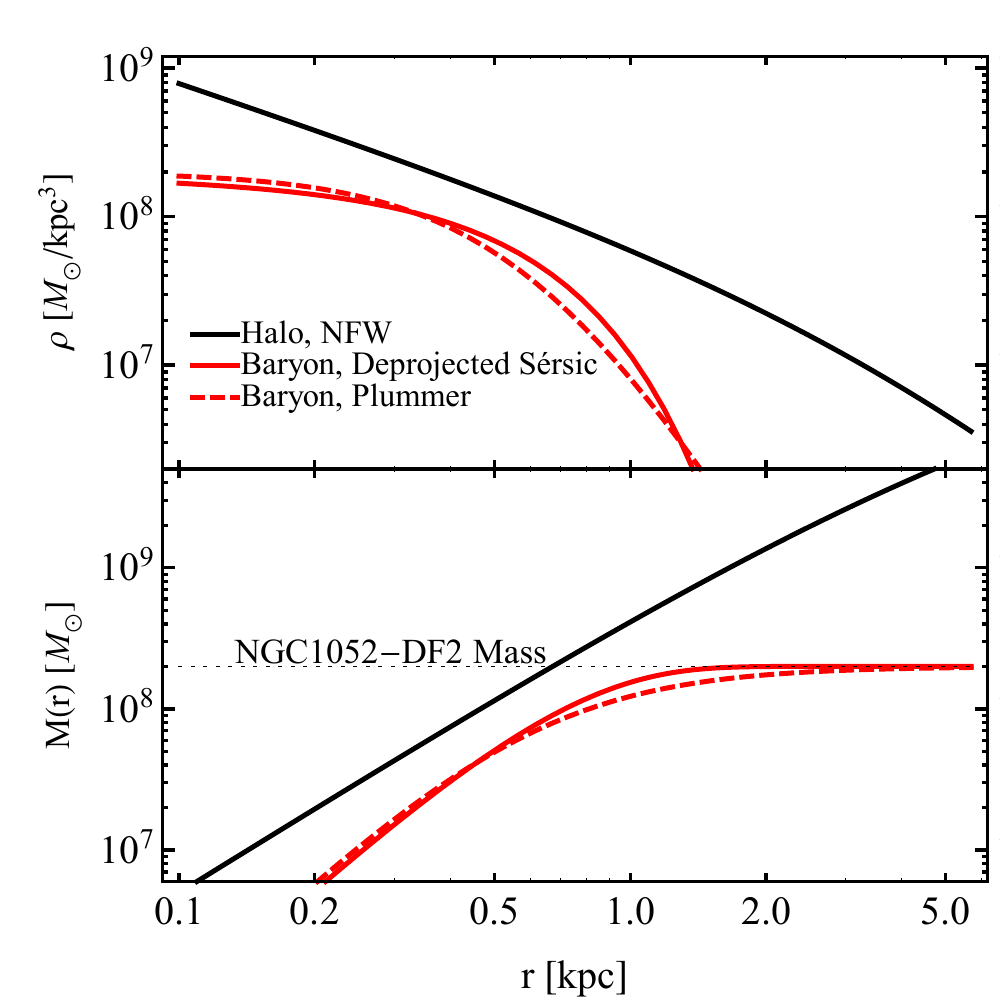}
\caption{\textbf{The simulation used initial density profiles for the halo $+$ star system of NGC 1052-DF2 (top panel), and the corresponding enclosed masses (bottom panel).} Here we assume the final NGC 1052-DF2 system is a linear stretch. Unlike NGC 1052, even in the very central region DM is still more than baryon, but an order a few DM to baryon ratio can be achieved.}
\label{fig:NGC1052DF2}
\end{figure}

When the NGC 1052-DF2 passes through the steep gravitational field of the central baryon cusp, tidal effect on it will significantly stretch the system in size, like the solar light is refracted to spread when passing through a prism. One can use the gravitational two body encounter problem solution~\cite{Binney&Tremaine} for a deflection estimation. For a single particle encounter
\begin{equation}
\theta=2\tan^{-1}\frac{G(M(b)+m)}{bv_\infty^2}\approx2\frac{GM(b)}{bv_\infty^2}\approx 2\frac{GM_\text{H}}{v_\infty^2}\frac{b}{(b+r_\text{H})^2},
\label{eq:deflection}
\end{equation}
where $b$ is the impact parameter, $M(b)$ is the enclosed mass at radius $b$ (here we neglect the difference of asymptotic $b$ with the true pericenter with gravitational bending) and in the last equation we have approximated with the Hernquist profile enclosed mass $M(r)=M_\text{H}r^2/(r+r_\text{H})^2$, $v_\infty$ is the asymptotic incident velocity and here by matching to N body simulation it is found to be the maximal velocity at pericenter $v_\text{peri}$. Different particles at different $b+\Delta b$ will be refracted differently, resulting a spread of the particles with angle difference $\Delta\theta\approx 2\frac{GM_\text{H}}{v_\text{peri}^2}\frac{d}{db}\big(\frac{b}{(b+r_\text{H})^2}\big)\Delta b$. Such $\Delta\theta$ introduces different normal kicks and is the source for heating the halo or star up, causing the stretch. In the inset of Fig.~\ref{fig:NGC1052} we plot such spread of $\theta$ with $b$. With the pericenter velocity $v_\text{peri}\sim700~\text{km}/\text{s}$ the introduced normal kick can be order of tens of $\text{km}/\text{s}$ at $\Delta b\sim\text{kpc}$, nonnegligible to the velocity dispersion determined by the profile. In this case after the refraction the observed $r=7.6$~kpc region can corresponds to a sufficiently inner region in the pre-encounter dwarf halo, so the enclosed mass can be sufficiently small in the inner region, consistent with the observed $M/L$ of order a few (see Fig.~\ref{fig:NGC1052DF2} the bottom panel).

In Fig.~\ref{fig:NGC1052DF2} we plot our test IC for NGC 1052-DF2. After the expected encounter the observation of NGC 1052-DF2 today gives an UDG with fitted parameter $M_\ast=2\times10^8~M_\odot$, $R_e=2.2$~kpc and S\'{e}rsic index $n=0.6$. In~\cite{vanDokkum:2018vup} a corresponding halo of $M_\text{halo}=6\times10^{10}~M_\odot$ (within virial radius it is $3.25\times10^{10}~M_\odot$) is suggested and we use it for the IC. The NFW profile parameter $r_s=5.69$~kpc is also the best match as above~\cite{Dutton:2014xda}. For realistic haloes it is well established that stellar feedback can reduce the central density and make it core-like, here for simplicity we still use the NFW one, and one should bear in mind that the final DM lacking effect can be more significant correspondingly. For a small S\'{e}rsic index the deprojected star profile is cored, so here we sample it with a Plummer profile which is conveniently provided in code, under the assumption that the stretch effect is linear. Ignoring small spherical asymmetry (ellipticity $b/a=0.85$) and deprojecting the above observed S\'{e}rsic profile gives the 3d half mass radius $r_\text{half}=2.88$~kpc and central density $\rho_c=3\times10^6M_\odot/\text{kpc}^3$, and matching it to that of a Plummer profile with the same total mass and central density gives $r_\text{P}=2.21$~kpc. However, the IC for $r_\text{P}$ before encounter cannot be determined since we do not know the stretch factor in advance. Realistic simulations~\cite{Hopkins:2013vha,2018MNRAS.480..800H} with feedback effect have well established that the baryon is relatively less concentrated as its larger scale counterpart (\emph{e.g.}, NGC 1052), and can develop core with kpc scales. Here we will use a baryonic central density of $\rho_c=2\times10^8M_\odot/\text{kpc}^3$, which is essentially from scan but roughly consistent with the early forming category of dwarf galaxy with such mass as in~\cite{2018MNRAS.480..800H}. In order to leave room for the refraction stretch this is a relatively small core, the larger ones can further reduce the central density by up to one order. Then the Plummer radius is $r_\text{P}=(3M_\text{H}/(4\pi \rho))^{1/3}=0.62~\text{kpc}$ and the half mass radius is $r_\text{half}=0.81~\text{kpc}$. In Fig.~\ref{fig:NGC1052DF2} in addition to the Plummer profile (dashed red curve) we also plot the ``shrinking'' of the final deprojected S\'{e}rsic profile (solid red curve), with the same total mass and the aforementioned central density for comparison. One can see that the Plummer profile is less flat and drops more quickly than the $n=0.6$ deprojected S\'{e}rsic profile around the radius of $r_\text{P}$.

We use N-body simulation code \texttt{Gadget2}~\cite{2005MNRAS.364.1105S} to study the encounter, in a variant of isolated simulation which enable us to use precisely controlled IC. The IC file is generated by a modified \texttt{SpherIC}~\cite{SpherIC} code, which sets two halo $+$ star system simultaneously. The NGC 1052 system and the NGC 1052-DF2 system are initially placed $400$~kpc apart, with a relative velocity of $250~\text{km}/\text{s}$ towards each other. This relative velocity is consistent with gravitational acceleration of free falling from infinity. In order to control the NGC 1052 stellar distribution precisely as inferred from observation which is crucial for our mechanism, we use a static baryonic potential kept to the center of NGC 1052 instead of live particles. The NGC 1052 halo and NGC 1052-DF2 halo are simulated by $5.54$ million and $0.06$ million halo particles respectively, and the NGC 1052-DF2 profile is simulated by $0.1$ million star particles. The gravitational softening length is $114$~pc for halo particle and $14.4$~pc for star particle.

\begin{figure*}[t!]
\includegraphics[height=3in]{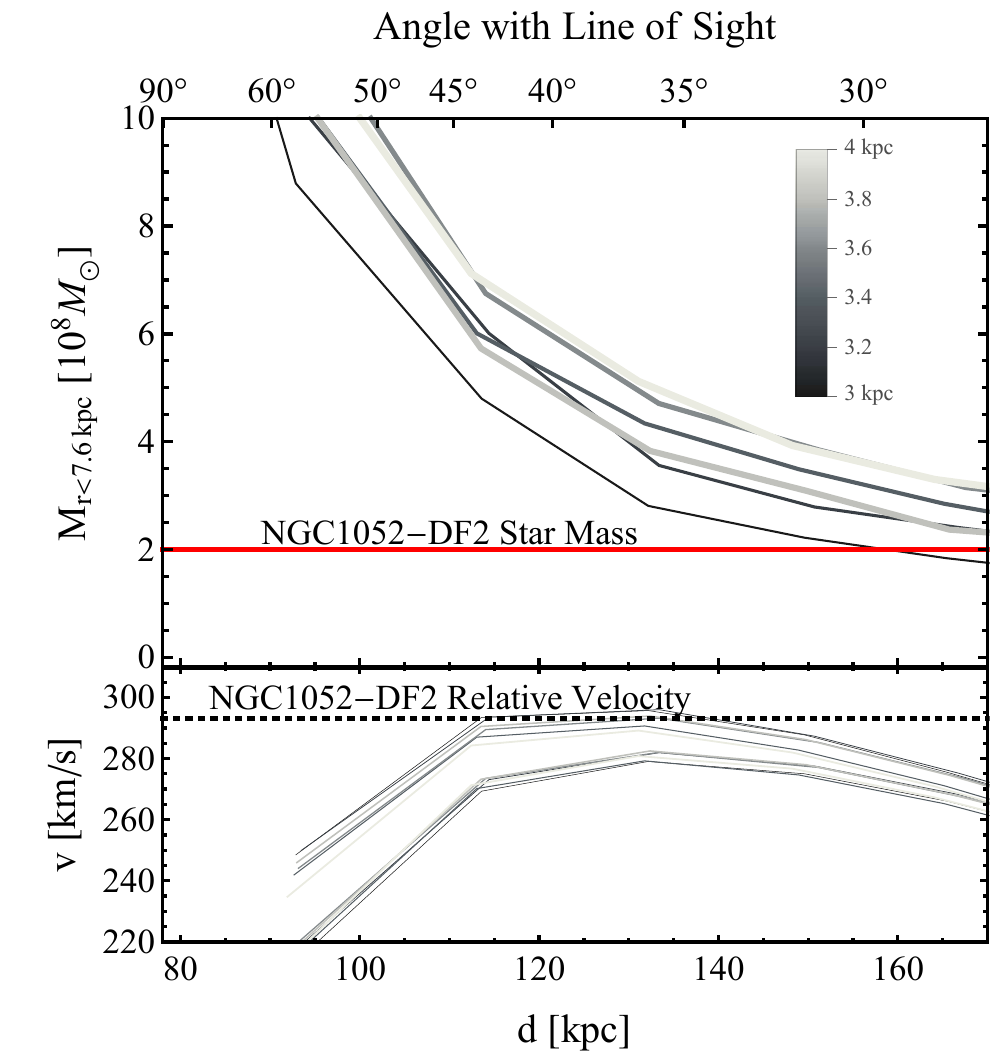}\qquad\includegraphics[height=2.73in]{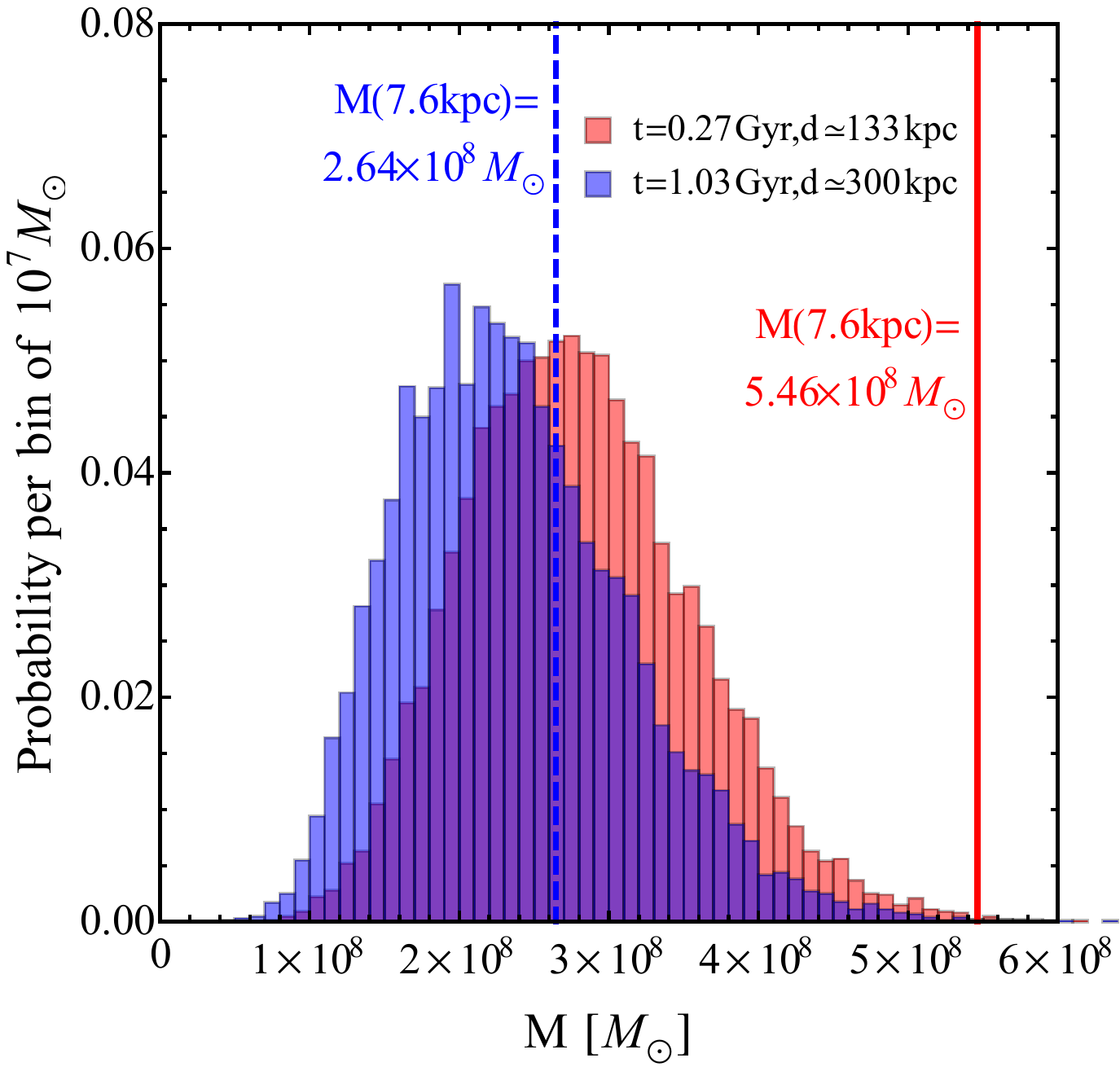}
\caption{\textbf{The combination of the two effects explains the observed lacking of DM in NGC 1052-DF2.} Left panel: DM enclosed within the $7.6~\text{kpc}$ from several simulation snapshots after encounter, in which we scan the impact parameter from $3.0$~kpc to $4.0$~kpc. In the bottom subpanel we also check the radial velocity relative to NGC 1052 and confirm that the observed $293~\text{km}/\text{s}$ can be reproduced within several snapshots. Right panel: The probability histogram by randomly selecting $10$ star velocities within the central $7.6~\text{kpc}$ and calculating their TME mass $M=C\langle\sigma^2r\rangle/G$, in comparison with the true enclosed mass from simulation. While after a sufficient long relaxation (blue histogram compared to blue vertical dashed line, the last snapshot of our simulation) the TME indeed converge to its true mass, in our interested snapshots with insufficient relaxation (red histogram compared to red vertical line) the TME tends to underestimate the enclosed mass systematically.}
\label{fig:halo}
\end{figure*}

In Fig.~\ref{fig:halo} we plot the information extracted from the snapshots which can be consistent with the observation after encounter, here and later $d$ is the true 3d distance between the centers of the two. First in the left panel we can see that with scan of impact parameter, all simulations give small enclosed DM mass within the $7.6$~kpc which can be consistent with the observation, at the level of $M/L<8$. This is a consequence of the initial radial distribution, that before encounter the enclosed mass at the corresponding radius has a similar ratio. We vary the impact parameter from $b=3.0$~kpc to $4.0$~kpc with a spacing of $0.2$~kpc, then we can see that generally the smaller the impact parameter, the larger the stretch and the smaller the enclosed mass in the $7.6~\text{kpc}$. In the lower subpanel we check the visual velocity relative to the central NGC 1052, and in a region it can be consistent with the observed value.

Apparently right after the encounter the system is experiencing substantial relaxation. Since the gravitational mass estimation relies on the effective dynamical equilibrium, deviation from such equilibrium can systematically affect the mass estimation. We have repeated the tracer mass estimator (TME) method~\cite{2010MNRAS.406..264W} as in~\cite{vanDokkum:2018vup,vanDokkum:2019}, by randomly selecting ten stars within the central $7.6$~kpc in the snapshots and calculating the enclosed mass estimation by $M=C\langle\sigma^2r\rangle/G$. We find it tends to give smaller mass estimations than the true mass during the interested snapshots systematically, which is plotted in the right panel of Fig.~\ref{fig:halo} and here $t$ is the time after encounter. This is the second reason accounting for the observed DM lacking in~\cite{vanDokkum:2018vup,vanDokkum:2019}.

\begin{figure}[t!]
\includegraphics[height=3in]{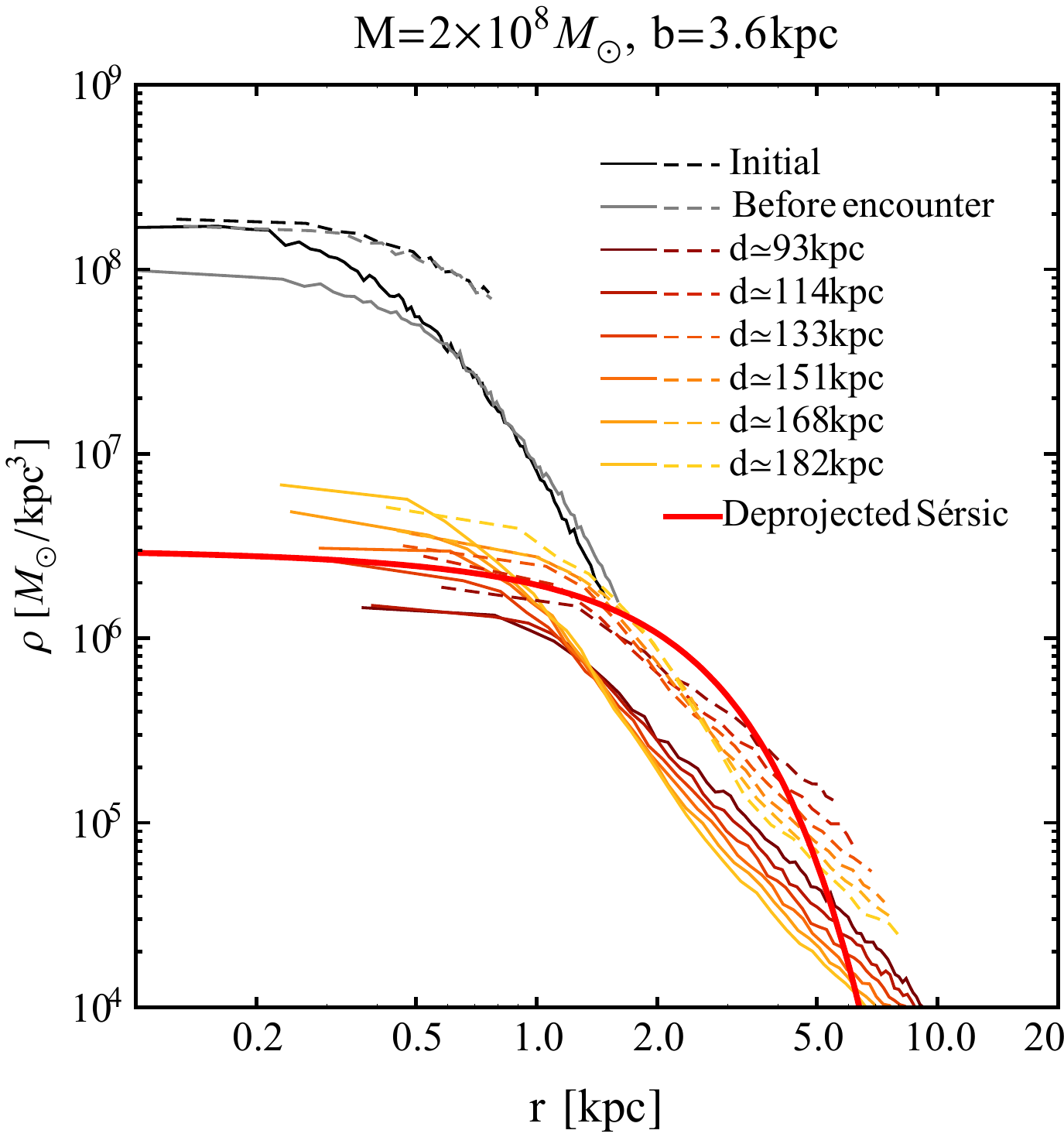}
\caption{\textbf{The baryonic profile extracted from several snapshots.} We assume spherical symmetry in making plot. With initial central density $\rho=2\times10^8~M_\odot/\text{kpc}^3$ and impact parameter $b=3.6~\text{kpc}$ we can reproduce the central density of the deprojected S\'{e}rsic profile. The cores from the initial Plummer profile (all solid curves) are smaller in size, and we compare them with the innermost $2\times10^8~M_\odot$ mass from a Plummer profile with the same central density $\rho=2\times10^8~M_\odot/\text{kpc}^3$ but larger in size $r_\text{P}=1.06~\text{kpc}$ (all dashed curves). A larger and more flat tuned initial core is able to reproduce the observed deprojected S\'{e}rsic profile.}
\label{fig:star}
\end{figure}

The observed star profile of the UDG as mentioned before can be viewed as a constraint that such process should reproduce. It is difficult for tidal stripping models, since the initial cusp set to prevent fast leakage of baryon is hard to get transformed into the observed core in evolution, and a tidal tail is usually left which breaks the neat shape as well. In Fig.~\ref{fig:star} we plot our star profiles. With our previously chosen central density $\rho_c=2\times10^8~M_\odot/\text{kpc}^3$ and correspondingly $r_\text{P}=0.62~\text{kpc}$ (solid curves), we find that $b=3.6$~kpc best reproduce the final deprojected S\'{e}rsic profile. We find the previous assumption that the stretch is linear acceptable. The anchor is that the central density $\rho_c=3\times10^6M_\odot/\text{kpc}^3$ is approximately reproduced. Using exactly such Plummer profile we cannot reproduce a large enough central core, which can be the consequence of the initial core not being large and flat enough. Keeping the same the other initial configurations such as $b$, we check by replacing the aforementioned Plummer profile by a larger Plummer profile with a larger total mass $M_\text{P}=10^9~M_\odot$ and larger radius $r_\text{P}=1.06$~kpc while keeping the central density fixed, and truncate to the innermost $2\times10^8~M_\odot$ mass to mimic a larger and more flat core (dashed curves). We can see that while the central density is kept, a larger core is reproduced which are closer to the observed deprojected S\'{e}rsic profile of the UDG. Given the freedom of tuning the initial star profile, we conclude that the observed UDG can be reproduced through such tidal stretch.

The assumption of such encounter agrees with the observational evidence that there is recent star burst in the NGC 1052~\cite{2011MNRAS.411L..21F}, where the impact of the dwarf's passage trigger such star burst. It also agrees with the gas lacking observation of NGC 1052-DF2~\cite{2019MNRAS.482L..99C,2019ApJ...871L..31S}, where gas is lost during the passage. In general we expect such close encounter to be rare, but the true rate estimation is beyond the current study.

Is there room of new physics in such scenario? One alternative to the cold DM model is if DM is self-scattering~\cite{Spergel:1999mh}, during passage the DM particles in the NGC 1052-DF2 halo may get scattered with the NGC 1052 main halo. And if the scattering is very efficient then NGC 1052-DF2 halo will be blocked by such scattering and not follow the UDG, opposite to the bullet cluster case in which it is baryon rather than DM that is partially blocked (\emph{e.g.,} we can arrange a velocity dependence of the self scattering cross section for this to be consistent with the bullet cluster). With this idea as the original motivation, we test such possibility with the same IC setup with a $\frac{\sigma_{\chi\chi}}{m_\chi}=2~\text{cm}^2/\text{g}$ DM self scattering cross section. We find that there is no noticeable difference with the cold DM simulation. This can be seen through an estimation of the expected count of DM scattering in a single passage
\begin{equation}
\langle N\rangle
=0.04\Big(\frac{\sigma/m}{2~\text{cm}^2/\text{g}}\Big)\Big(\frac{\rho}{10^7~M_\odot/\text{kpc}^3}\Big)\Big(\frac{d}{10~\text{kpc}}\Big)
\label{eq:SIDMnumber}
\end{equation}
The expected number is much smaller than one, meaning the scattering will not be significant to cause a block effect.

\textbf{Method}

\textbf{Additional Information of Initial Profile.} For baryonic profile assuming spherical symmetry, since the S\'{e}rsic profile $I(R)$ is get from a 3d profile $\rho(r)$ through a projection onto 2d plane $I(R)=2\int_R^\infty d\sqrt{r^2-R^2}\rho(r)=2\int_R^\infty \frac{rdr}{\sqrt{r^2-R^2}}\rho(r)$, the deprojection from S\'{e}rsic profile back into 3d is done by an inverse Abel transformation
\begin{equation}
\rho(r)=-\frac{1}{\pi}\int_r^\infty\frac{dr'}{\sqrt{r'^2-r^2}}\frac{dI(r')}{dr'}.
\label{eq:deprojection}
\end{equation}

The Hernquist profile $\rho(r)=\rho_0/\big(\frac{r}{r_\text{H}}(1+\frac{r}{r_\text{H}})^3\big)$ gives a finite total mass asymptotically, for NGC 1052 it is $M_\text{H}=M_\ast=10^{11}~M_\odot$, and enclosed mass $M(r)=\frac{r^2}{(r+r_\text{H})^2}M_\text{H}$. The other parameter $r_\text{H}$ can be determined by matching the half mass radius to the deprojected S\'{e}rsic profile, the 3d half mass radius of deprojected S\'{e}rsic profile is $2.83$~kpc and the Hernquist profile has $r_\text{half}=(1+\sqrt{2})r_\text{H}$, so $r_\text{H}=1.17$~kpc. Similarly the Plummer profile $\rho(r)=\rho_c/\big(1+(\frac{r}{r_\text{P}})^2\big)^{5/2}$ also gives a finite total mass, for NGC 1052-DF2 it is $M_\text{P}=M_\ast=2\times10^8~M_\odot$. The Plummer profile enclosed mass is $M(r)=\frac{r^3}{(r^2+r_\text{P}^2)^{3/2}}M_\text{P}$. The central density can be determined as $\rho_c=M_\text{P}/(\frac{4}{3}\pi r_\text{P}^3)$.

On the other hand, the NFW profile concentration $-$ mass relation of~\cite{Dutton:2014xda} is $\log_{10}c_{200}=0.905-0.101\log_{10}(M_{200}/(10^{12}M_\odot/h))$ or $\log_{10}c_\text{vir}=1.025-0.097\log_{10}(M_\text{vir}/(10^{12}M_\odot/h))$, here we use the former while the difference with the latter should be small. The NFW profile diverges logarithmically in total enclosed mass at large $r$, so the initial condition (IC) generator \texttt{SpherIC} use a smooth transition from the inner NFW profile to the outer exponential decay behavior at some cutoff radius for an NFW profile, here globally we take $r_\text{cutoff}=7.84r_s$ and have ignored the size dependence. In that case the IC generator will generator $71.5\%$ halo particles within the cutoff radius. The NGC 1052 NFW halo has $\rho_0=3.64\times10^6M_\odot/\text{kpc}^3$, and the NGC 1052-DF2 halo has $\rho_0=1.43\times10^7M_\odot/\text{kpc}^3$.


\textbf{Modification of \texttt{SpherIC} and \texttt{Gadget2}.} We modify the IC generator \texttt{SpherIC} to allow two halo $+$ star system generated simultaneously, basically by repeating the generation of a single overlapping halo $+$ star system. For each halo $+$ star system an overall displacement and overall velocity can be set respectively in the same way as the original \texttt{SpherIC}, which gives very precise control of the IC. The halo particle in the two system share the same mass and so does the star particle, which are required by the simulation code \texttt{Gadget2}.

As aforementioned we do not sample the star of NGC 1052 by living particles, instead we add in \texttt{Gadget2} a static potential which is always overlapping with the center of the halo, by checking the center (average position in each dimension) at each time step. The baryonic potential is obtained by numerical integration of the deprojected cored S\'{e}rsic profile, see equation (\ref{eq:deprojection}). Such implementation will not be faithful to using living particles since it ignores individual close encounter, but taking such effect into account is expected to enhance the spreading of the particles refraction, and making result less conservative. On the other hand, to practical simulation resolution the star particle mass will be several orders larger than a true star mass, and the close encounter effect will be too larger than it truly can be if using living particle.

\textbf{Parameter Scan and Determination.} Before finalizing the parameter with reference to the baryonic simulations we have performed several round of scan of relevant parameters, \emph{e.g.}, scan of impact parameter $b$ from $2.0$~kpc to $4.0$~kpc with spacing of $0.2$~kpc, or scan of the Plummer radius from $0.2$~kpc to $0.6$~kpc with total Plummer mass fixed to a larger value $4\times10^8~M_\odot$. We have found that for small Plummer radius so that the central density $\rho_\text{P}\gtrsim 4\times10^8~M_\odot/\text{kpc}^3$, the tidal impulse is insufficient to stretch the baryon to produce a sufficient large and flat core as the observed deprojected S\'{e}rsic profile, and after stretch the central region of the NGC 1052 tends to contract back already significantly during the interested snapshots, leading to a small dense core. And on the other hand, the larger the Plummer radius the smaller the stretch effect is according to the final ultra-diffuse galaxy size, and the more the enclosed dark matter correspondingly. So we determine the optimistic central density to be $\rho_\text{P}= 2\times10^8~M_\odot/\text{kpc}^3$ and the Plummer radius $r_\text{P}=0.62$~kpc. From various scan we see that the final central density of star core is determined by the initial central density and impact parameter, but almost irrelevant to the initial core size.

We expect a neighborhood of our best parameter point also gives good enough DM and ultra-diffuse galaxy phenomena reproduction as observation.


\textbf{Visual Geometry.} A typical deflection angle of the trajectory at $b=3.6$~kpc is $\theta=0.2$ (from simulation). We do not at first fix the relative direction of the trajectory of the encounter toward us which is $20$~Mpc away from the very beginning, but we choose several snapshots to check what relative angle can reproduce the observed $80$~kpc projected distance. In simulations with snapshot frequency of every $0.05$~Gyr, the snapshots are the $3$rd to $9$th ones after the encounter, or the $23$rd to $29$th totally from the very beginning of $400$~kpc away. The angle of the NGC 1052 to NGC 1052-DF2 connection direction with the line of sight is ticked on the top side of Fig.~\ref{fig:halo}. On the other hand, the NGC 1052 to NGC 1052-DF2 connection direction is not exactly coinciding with the relative velocity direction of NGC 1052-DF2 but has a small angle, since it is not exactly passing the center of NGC 1052. With line of sight on the two sides of the trajectory we have two corrections and two group of projected velocity on the line of sight. With our parameter choice we find the line of sight at the opposite side of the deflection gives higher line of sight velocity and agrees better with the observed $293~\text{km}/\text{s}$ line of sight relative velocity. It is just a demonstration and we expect such configuration can still be reproduced with other parameter choices.

\textbf{Tracer Mass Estimator.} The tracer mass estimation mass is $M(r)=C\langle\sigma^2r\rangle/G$~\cite{2010MNRAS.406..264W}. Here
\begin{equation}
C=\frac{4\Gamma(\frac{\alpha}{2}+\frac{5}{2})}{\sqrt{\pi}\Gamma(\frac{\alpha}{2}+1)}\frac{\alpha+\gamma+1-2\beta}{\alpha+3-\beta(\alpha+2)}
\end{equation}
is an order one parameter, where $\alpha$ controls the slope of the potential of the total enclosed mass ($\rho\propto r^{-(\alpha+2)}$), $\beta=1-\sigma_t^2/\sigma_r^2$ is the Binney anisotropy parameter~\cite{Binney&Tremaine}, and $\gamma=0.9\pm0.3$ is the potential of the tracer (global cluster) profile itself. Here we take $\alpha=0$ and $\beta=0$ as in~\cite{vanDokkum:2018vup}, then $C=1.9$. From simulation we know the velocities of the selected star particles precisely, so we use the root mean square velocity dispersion and average over 3d to find the 1d value (consistent with $\beta=0$). For simplicity we consistently use the true 3d largest radius of the selected star particles. We also checked such method gives good estimation for the NGC 1052. The Fig.~\ref{fig:halo} right panel histogram are based on statistics of $10^4$ trials for each case.

\textbf{Spherical Asymmetric Distortion.} Using the kernel density estimation along the three dimensions we have checked that the shape of the star profile after the encounter at the interested snapshots have no significant deviation from spherical symmetry, especially in the inner region. The very outskirt of the post-encounter star profile tends to slightly lag in the direction towards the NGC 1052 and against the bending direction, but no baryonic tidal tail is observed. Freedom in spherical asymmetry in IC also provides room to account for the existence or nonexistence of shape distortion.

\bibliography{NGC}

\textbf{Acknowledgements.} We acknowledge the computational facilities of the High-Performance Computing Center at UC Riverside and the chepfarm cluster at Tsinghua University.

\end{document}